\documentclass[12pt,preprint]{emulateapj}

\newcommand{\drizzle}{{\sc drizzle}}
\newcommand{\iraf}{{\sc iraf}}

\def\lsim{\mathrel{\rlap{\lower4pt\hbox{\hskip1pt$\sim$}}
    \raise1pt\hbox{$<$}}}                % less than or approx. symbol
\def\gsim{\mathrel{\rlap{\lower4pt\hbox{\hskip1pt$\sim$}}
    \raise1pt\hbox{$>$}}}                % greater than or approx. symbol

\def\swift{{\it Swift}}

\shorttitle{Late time observations of GRB\,080319B}
\shortauthors{Tanvir et al.}

\begin{document}


\title{Late time observations of GRB\,080319B: jet break, host galaxy and accompanying supernova}


\author{N. R. Tanvir}
\affil{Department of Physics and Astronomy, University of Leicester,
University Road, Leicester, LE1 7RH, United Kingdom}
\email{nrt3@star.le.ac.uk}

\author{E. Rol}
\affil{Astronomical Institute ``Anton Pannekoek'', P.O. Box 94248, 1090 SJ Amsterdam, The Netherlands}

\author{A. J. Levan, K. Svensson}
\affil{Department of Physics, University of Warwick, Coventry, CV4 7AL, United Kingdom}

\author{A. S. Fruchter}
\affil{Space Telescope Science Institute, 3700 San Martin Drive, Baltimore, MD~21218, USA}

\author{J. Granot}
\affil{Centre for Astrophysics Research, University of Hertfordshire, College Lane, Hatfield, AL10 9AB, United Kingdom}

\author{P. T. O'Brien, K. Wiersema, R. L. C. Starling}
\affil{Department of Physics and Astronomy, University of Leicester,
University Road, Leicester, LE1 7RH, United Kingdom}

\author{P. Jakobsson}
\affil{
Centre for Astrophysics and Cosmology, Science Institute, University of Iceland, 
Dunhagi 5, IS 107 Reykjavik, Iceland}

\author{J. Fynbo, J. Hjorth}
\affil{Dark Cosmology Centre, Niels Bohr Institute, University of Copenhagen, Juliane Maries vej 30, 2100, Copenhagen, Denmark}

\author{P. A. Curran}
\affil{AIM, CEA/DSM - CNRS, Irfu/SAP, Centre de Saclay, Bat. 709, FR-91191
Gif-sur-Yvette Cedex, France}

\author{A. J. van der Horst\footnote{NASA Postdoctoral Program Fellow}, C. Kouveliotou}
\affil{NASA/Marshall Space Flight Center, NSSTC, 320 Sparkman Drive, Huntsville, Alabama 35805, USA}

\author{J. L. Racusin}
\affil{NASA Goddard Space Flight Center, 8800 Greenbelt Rd., Code 661, Greenbelt, MD 20771, USA}

\author{D. N. Burrows}
\affil{Department of Astronomy and Astrophysics, 525 Davey Lab, Pennsylvania State University,
University Park, PA, USA}

\and

\author{F. Genet}
\affil{Racah Institute of Physics, Hebrew University, 91904 Jerusalem, Israel}
%Centre for Astrophysics Research, University of Hertfordshire, College Lane, Hatfield, AL10 9AB, United Kingdom


%\altaffiltext{2}{Society of Fellows, Harvard University.}


\begin{abstract}
The \swift\ discovered GRB\,080319B was by far the most distant source ever observed at naked eye 
brightness, reaching a peak apparent magnitude of 5.3 at a redshift of $z=0.937$.  
We present our late-time optical ({\it HST}, Gemini \& VLT)  and X-ray ({\it Chandra}) observations, which 
confirm that 
an achromatic break occurred in the power-law afterglow light curve  at 
$\sim11$\,days
post-burst.  
This most likely indicates that the gamma-ray burst (GRB) outflow was collimated,
which
for a uniform jet would imply
a total energy in the jet $E_{\rm jet} \gsim 10^{52}\;$erg.
Our observations also show a late-time excess of red light, 
which is well explained if
the GRB was accompanied
by a  supernova (SN), similar to those seen in some other long-duration GRBs.
The latest observations are dominated by light from the host and
show that the GRB took place
in a faint dwarf galaxy ($r({\rm AB})\approx27.0$, rest-frame $M_B\approx-17.2$).
This galaxy is small even by the standards of other GRB
hosts, which is suggestive of a low metallicity environment.
Intriguingly, the properties of this extreme event -- a small host and bright supernova --
are entirely typical of the very low-luminosity bursts such as GRB\,980425 and GRB\,060218.
 \end{abstract}


\keywords{gamma ray bursts:individual (GRB080319B) - galaxies: high-redshift - supernovae: individual}



\section{Introduction}

GRB\,080319B was one of the brightest gamma-ray bursts (GRBs) yet seen
in gamma-rays, and uniquely bright in optical and X-ray wavelengths.
At a redshift of $z=0.937$ \citep{vreeswijk08} this also translates to a
record-breaking intrinsic peak  luminosity in the optical, being approximately 2 magnitudes brighter than
GRB 990123 \citep{akerlof99} 
and a magnitude brighter than GRB 050904 \citep{haislip06}.

By good fortune, an earlier burst, GRB 080319A, had already taken
place nearby on the sky roughly 25 minutes before GRB\,080319B, so several wide-field optical cameras
obtained imaging of the prompt phase, giving unprecedented
coverage of the optical flash, and showing it to reach a visual
magnitude 5.3 \citep{racusin08}.

Despite (or perhaps because of) the exceptionally dense 
multi-wavelength coverage of this
event and its afterglow, modeling its properties has proven difficult.  
A number of authors initially argued that the
(soft) gamma-ray component was likely
dominated by synchrotron self-Compton (SSC), i.e. inverse-Compton
up-scattering of (optical) synchrotron photons that are produced by the
same population of relativistic electrons.  This was supported by 
rough similarity of the optical and gamma-ray prompt light curves.
If this were the case, then
2nd-order SSC should create another peak of emission in the GeV
regime, of even greater total fluence \citep{kumar08,racusin08,piran09}.  This potentially leads to a
serious energy crisis, with the total radiated and kinetic energies,
if isotropic, being comparable to or even in excess of the rest-mass energy of a massive
star.  

Subsequent analyses have been unable to construct a consistent
SSC model, and have argued instead that the two (optical and soft
gamma-ray) prompt components must be produced in different regions
\citep{Zou09} or that they are produced by a relativistically turbulent
outflow, rather than internal shocks, at relatively large radius 
\citep{Kumar09}.

The later time behaviour has proven similarly contentious.
It has long been thought that GRB outflows are likely to be collimated
into narrow jets, 
and that this could reduce the total energy requirement 
by 1--3
(and in extreme cases perhaps more)
orders of magnitude.
The observational signature of such beaming
is an achromatic  break (hereafter referred to as a ``jet-break")
in the power-law decline of afterglow light \citep{rhoads99,sari99}.
However, the luminosity of GRB\,080319B and its afterglow
may still stretch plausible models for both the
prompt and afterglow emission.

\citet{racusin08} proposed a model in which the jet giving rise to the
GRB has a particularly high-velocity, bright and narrow ($\sim0.2^{\circ}$) core
which produces a jet-break $\sim1$ hour post burst, and dominates the early emission.  A wider
($\sim4^{\circ}$), more
``conventional" jet surrounds this and 
dominates at intermediate and late times.
This second jet is assumed to give rise to the 
break at $\sim 10^6$ seconds seen in the {\em Swift}/XRT light curve.

In a model of this sort, the extreme behaviour of the burst is
partially explained by the low probability of an observer being within
the aperture of the narrow jet. For GRB\,080319B the fraction of
observers viewing the gamma-ray emission from the 
bright and narrow jet would be
roughly a factor of 400 lower than those seeing the broad jet.
It also provides a reasonably good description of aspects of the
temporal evolution of the afterglow.
However, the model also requires a further
coincidence of a (rarely seen) strong reverse shock from the wider jet creating the
early optical afterglow, and this double coincidence seems a less natural scenario.
We also note that such an
extreme ratio of opening angles and solid angles between the wide and
narrow jet components is much larger than the ratio of $\sim 3$ in
opening angles expected in the original motivations for the two component
jet models, which include the cocoon in the context of the collapsar
model and the neutron decoupling during the acceleration and collimation
of a hydromagnetic jet \citep[see][and references therein]{peng05}. 
Furthermore, the required half-opening angle of the narrow jet
($0.2^\circ$) is extremely small, and only slightly above the inverse of
the initial Lorentz factor.

An alternative model developed by \citet{racusin08} has a single
jet, ploughing into a complex density medium.  In this case,
the evolution of the cooling break frequency is proposed to drive the
changes in the broad spectral-energy distribution (SED) of the afterglow.

Regardless of the successes and limitations of such models,
it is clearly of great interest to investigate the late-time behavior of
GRB\,080319B and to place it in context with other bursts, which may
provide independent clues to its nature. 
Is the late time evolution 
comparable to that seen in most long duration GRBs? In particular, 
is the sharp break in the X-ray lightcurve at $\sim10^6$ seconds achromatic,
as predicted for a jet break, and what does
this imply for the energetics of the burst?  Is the burst accompanied
by a characteristic type Ic supernova (SN)?  
Is the underlying host galaxy similar
to those of other long-duration GRBs?

In this paper we describe our late-time optical and X-ray
monitoring of the transient and host
galaxy emission of GRB\,080319B, utilizing Gemini-North, {\em HST}, the VLT and 
{\em Chandra}.

%%%%%%%%%%%%%%%%%%%%%%%%%%%%%%%%%%%%%%%%%%%%

\section{Observations and reduction}

\subsection{X-ray observations}

In order to follow the X-ray lightcurve out to late times, beyond the
sensitivity limit of the {\it Swift}/XRT, we obtained observations with
{\it Chandra}/ACIS (S3 chip), roughly 38 and 58 days after the burst. 
We used the standard processed data (ASCDS version 7.6.11.6) for our
analysis, selecting an energy range between 0.3 and 7\,keV, which gave
an optimal signal to noise. Photometry was performed with a 5 pixel (2.5\,arcsec) radius
region centred at the source position, and an
annular region centred around the source as the background region
(inner radius 14\,arcsec, outer radius 28\,arcsec).

The first epoch consisted of a 9\,ks exposure, with 9 counts detected
in the source region and a predicted background of 0.4 counts.
The second epoch was a 36\,ks exposure, resulting in 18 counts in
the source region and a 1.48 count background.
Data were fit inside XSpec using appropriate response matrices, with
the actual fitted values for photon index and absorption
from the {\it Swift}/XRT late time data \citep{racusin08}; 
thus, only the normalisation was fit. The fluxes were then
derived using this normalized fit in the 0.3 -- 10 keV range, giving 
absorption corrected values of
$8.4_{-2.7}^{+3.2}\times 10^{-15}\,{\rm erg~cm}^{-2}~{\rm s}^{-1}$ and
$3.7_{-0.8}^{+1.0}\times 10^{-15}\,{\rm erg~cm}^{-2}~{\rm s}^{-1}$ 
for epochs 1 and 2, respectively. 
Here, the $1\sigma$ errors were derived using
Bayesian confidence
limit estimation (Kraft et al. 1991).

\subsection{Optical/nIR observations}
Due largely to its brightness, early optical and near-infrared (nIR) observations of GRB\,080319B were
pursued by several groups, resulting in a very well-sampled optical/nIR lightcurve
covering the first few hours after the burst 
\citep{racusin08,bloom09,wozniak09,pandey09}.
Despite its initial brightness the afterglow faded rapidly, and photometric monitoring 
required large aperture telescopes after a few days. 

A log of all our late-time observations 
is provided in Table~\ref{tbl-1}. 
This does not include any correction for dust extinction:
the foreground extinction is expected to be small
\citep[$A_V=0.037$,][]{schlegel98} while the extinction internal
to the host, although rather uncertain due to the presence of a 
break between the optical and X-ray, is also found to be modest \citep{racusin08}.

We obtained optical observations with Gemini-N/GMOS, VLT/FORS1 and
{\em HST}/WFPC2, between $\sim3$ and $\sim460$ days post-burst.  
Processing of ground-based observations 
was performed using standard \iraf\ routines.  In particular
the GMOS reduction made use of the relevant customized software provided
by Gemini.
Photometric calibration, both zero-point and color terms, were obtained using Sloan Digital Sky Survey
(SDSS) stars in the field \citep{sdss,cool08}.
For consistency, the FORS1 $B$-band imaging was also calibrated to AB magnitudes.

For our {\em HST}/WFPC2 observations we placed the target on the WFALL 
aperture, on the corner of WFC3 closest to the apex, in order to 
reduce the impact of charge transfer inefficiency (CTE effect), which is
significant for the old detectors operating on WFPC2. 
A 4-point
dither pattern was used, and subexposures stacked 
using the \drizzle\  \citep{fruchter02}
software onto a 0.05 arcsec pixel grid (from the native 0.1 arcsec pixels).
Photometry of the transient was obtained in a 0.2 arcsec diameter aperture,
and aperture corrections to the standard 1 arcsec diameter calculated using brighter
point sources on the frame.
CTE correction was performed using the method of Dolphin\footnote{http://purcell.as.arizona.edu/wfpc2\_calib},
although we applied only half the correction to the final epoch since the source is clearly
extended\footnote{\tiny http://www.stsci.edu/hst/wfpc2/documents/isr/wfpc2\_isr0004.html}
(extra allowance was made in the error budget for this step).
The {\it HST} photometry was calibrated to AB magnitudes 
by reference to the tabulated zero points\footnote{\tiny http://www.stsci.edu/documents/dhb/web/c32\_wfpc2dataanal.fm1.html},
and then transformed to SDSS r and i magnitudes for comparison with
the ground-based data via the NICMOS Unit Conversion Form\footnote{http://www.stsci.edu/hst/nicmos/tools/conversion\_form.html}.

The position of the afterglow was determined, relative to two well
positioned 
SDSS stars in the field to be: RA=14:31:40.994,
dec=+36:18:08.64 (J2000), with an error of 0.02 arcsec in each coordinate.\footnote{Specifically the comparison
stars used from SDSS release 6 were
587736943056454244 at RA=14:31:41.866, dec= +36:17:23.13
and 587736943056453784 at RA=14:31:42.912, dec= +36:18:24.26.}

%%%%%%%%%%%%%%%%%%%%%%%%%%%%%%%%%%%%%%%%%%%%

\section{Results}

Figure 1 shows the summed {\it HST}/WFPC2 images at the three epochs of
observation.  The afterglow luminosity clearly declines with time,
being ultimately dominated by light from the host.
Our new X-ray and optical photometry is plotted in Figure~\ref{figlc},
together with data from the literature.
We expect the optical light curve to consist of three components:
the afterglow of the GRB, any accompanying supernova and
a steady underlying host galaxy.  In the X-ray the emission is
likely to be entirely from the afterglow.

In disentangling these components, 
our approach is to 
compare the photometry 
(corrected for
the small foreground Galactic extinction)
with a simple,
self-consistent model of a power-law
afterglow with a sharp achromatic break, and supernova.
As we will show, this model matches the broad features
of the data well, and allows us to focus on the main
implications of the late-time observations, without getting
embroiled in the fine details of the earlier time evolution.

We determine the afterglow power-law slopes solely from the X-ray light 
curve (adopting the convention flux
$F\propto t^{-\alpha}\nu^{-\beta}$),
finding 
$\alpha_1=1.28\pm0.04$,
characteristic of the pre-break decline between
$3\times10^4$\,s and $5\times10^5$\,s, and $\alpha_2=2.33\pm0.37$
characteristic of post-break between $1.2\times10^6$\,s and $4\times10^6$\,s.  
The break is taken to be abrupt and we find a best fit at
$11.6\pm1.0$\,days, whilst the spectral slope through the optical
bands is taken as $\beta=0.5$ \citep{racusin08}.  
We note that these values satisfy the closure relations for 
expansion into a wind-like medium when the cooling break is
situated above the optical \citep{Price02}.

Finally, the supernova light curve is based 
on that of SN1998bw,
but faded by 0.3 mags
consistent
with what is found for several other GRB-SNe \citep{galama00,zeh04}.
We also include {\em in the model} a small amount of rest-frame 
extinction internal to the host 
of $A_V=0.06$ \citep{wozniak09},
and assume a Small Magellanic Cloud (SMC) extinction law \citet{pei92} .

We expect
the shorter wavelength observations ($B$- and $g$-band)
to be largely uncontaminated by supernova light,
since SNe are weak rest-frame blue and UV emitters due to metal
line blanketing. The redder bands therefore constrain
any supernova component, which should
rise to a peak roughly one month post-burst. 
Finally our latest time observations
are dominated by host galaxy emission.  
However, as shown below, even at 
a few hundred days post-burst the photometry is still likely to 
be contaminated by some residual transient light.  
To allow for this we adopted an iterative procedure: first assuming the
last epoch  
shows only host,
hence using this magnitude to correct the earlier photometry for host 
contribution, and thus allowing modeling of the transient emission.
From this model we can then predict the remaining transient 
flux which is likely to still be in the latest observation in each band,
and hence we can re-estimate the host magnitudes with this
contamination removed.
The correction is about
10\% in $g$, and only a few percent in $r$ and $i$,
and we have
increased the photometric error bars to allow that the contamination
could actually range from zero up to this value.  It is worth
emphasizing
that the correction even in $g$ 
only has a small effect on all
but the $50+$ day photometric points, and does not change the 
main conclusions we present.

For the {\it HST} observations, since a much smaller aperture can be used,
the contribution of host light is less.  Here we make the maximally conservative
assumption that $0.5\pm0.5$ of the flux
measured in the final {\it HST} epochs is transient light, and this is then used to
correct the earlier epochs for host contamination.

\subsection{Jet-break and energetics}

Figure~\ref{figlc} (top left panel) shows our late time {\em Chandra} observations, as well 
as early data taken by the {\em Swift}/XRT. Our X-ray observations confirm, 
and increase the confidence in, the break in the X-ray lightcurve at 
$t_b \approx11$\,days.

The photometry for the various optical bands, with host 
contribution subtracted (as described above; see also Section 3.3), 
is plotted in the other panels
of Figure~\ref{figlc}.
In the $B$ and $g$-band observations, the light curve before and
after the break is reasonably
consistent with the X-ray slope and break time, indicating
approximately achromatic behaviour, as expected
for a jet break.
In fact, this is one of the more convincing examples of a jet-break 
identified in the {\em Swift} era, when such clear achromatic behavior
of X-ray and optical light curves
has rarely been seen \citep[e.g.][]{curran08}.

Since the jet-break time we find is consistent with that used in 
earlier studies, notably \citet{racusin08} and \citet{bloom09}, 
those analyses, and in particular their discussions of deviations
from a simple power-law at earlier times, are not modified by
our findings.

It is instructive to consider the simple case in which 
the break 
is interpreted in the context of
a single jet (double sided, roughly uniform with
reasonably sharp edges). Then a break time, $t_b\approx11\,$days, implies a half-opening angle of $\theta_{j}
\sim 10^\circ$, for a canonical external medium density of $n \sim 1\;{\rm
cm^{-3}}$ and (isotropic equivalent) kinetic energy comparable to
the energy observed in gamma-rays, $E_{k,\rm iso} \sim E_{\rm\gamma,iso} =
1.4\times 10^{54}\;$erg \citep{sari99}. This, in turn, implies a true energy output in
gamma-rays within the observed energy range of 
$E_\gamma \sim 2 \times10^{52}\;$erg, and a comparable kinetic energy in the jet ($E_{k}
\sim E_\gamma$).

Alternatively, the kinetic energy can also be estimated from the X-ray luminosity at 
$12\;$hr in the rest frame \citep{granot06,nousek06},  
from which we find $E_{ k,\rm iso}\approx 7\times10^{52}\;$erg,
for typical microphysical parameters 
($\epsilon_e=0.1$, $\epsilon_B=0.01$, $p=2.2$).
This corresponds to $\eta\equiv E_{ k,\rm iso}/E_{\gamma,\rm iso} \approx 0.05$ and
would in turn imply $\theta_{\rm j}
\sim 12^\circ$ and a true kinetic energy of 
$E_{ k} \sim 2 \times 10^{51}\;$erg.
The isotropic equivalent kinetic energy at this level would require a 
very high efficiency of the gamma-ray emission ($\gtrsim 95\%$)  for a single wide jet,
unless the microphysical parameters were very different so that
$E_{k,\rm iso}$ would be significantly higher.
If, on the other hand, the gamma-rays were produced by a narrow jet with a 
considerably higher $E_{k,\rm iso}$ then this can bring down the efficiency requirements to more reasonable values 
\citep{peng05}.
In fact, this feature is built into
the two component jet model of \citet{racusin08}, which postulates a very
narrow ($\theta_{j,n} \sim 0.2^\circ$),  very high Lorentz factor ($\Gamma\sim10^3$)
central jet, producing an early  break in the light curve, coupled with a wider
($\theta_{j,w} \sim 4^\circ$) jet leading to the later time break
we see at $\sim10^6\;$s. 
This model also
mitigates the energy crisis more effectively, with each jet producing $E_{\gamma}\approx2\times10^{50}\;$erg.

Finally, we draw attention  to the sharpness of the late-time
jet-break
as seen in the X-rays, which is also consistent with the optical observations,
notably in the $g$-band.  Such a sharp break is not expected
for a wind-like external medium \citep{kumar00}, as considered by \citet{kumar08},
and so would require some modification to that simple model, which otherwise nicely fits
the afterglow data between $\sim10^3$ and $\sim10^6\,$s.
One possibility would be  the coincidental
presence of a wind-termination shock in the ambient medium surrounding
the progenitor, at approximately 
the same radius at which the jet break occurs\footnote{Note that no sharp bump is expected in the lightcurve
when the afterglow shock encounters the wind termination shock\citep{Nakar07}}.
If we again consider a simple wide jet, this radius is given 
by $R_j = 1.2\times10^{19}(E_{k}/10^{51})(A_*/0.03)^{-1}\rm\, cm$,
where $A_*=(\dot{M}/10^{-5}\,{\rm M_{\odot}\,yr^{-1}})/(v/10^8\,{\rm cm\,s^{-1}} )$ 
is the conventional mass-loss scaling \citep[cf.][]{Panaitescu2000}.
Using
the relations $E_k = \eta E_{\gamma,{\rm iso}}\theta_j^2/2$ and $\theta_j^2 =
[16\pi Ac^3 t_b/(1+z)E_{\gamma,{\rm iso}}\eta]^{1/2}$ together with the
observed values of $z$, $E_{\gamma,{\rm iso}}$ and $t_b$  gives
$R_j = 3.1\times10^{19}\eta^{1/2}(A_*/0.03)^{-1/2}\rm\, cm$.
\citet{racusin08} argued for a tenuous wind with an upper limit on $A_*<0.03$
and $\eta\sim0.07$.

Now we obtain the wind termination shock radius 
using Equation (3) of \citet{Peer2006}:
$R_0 = 9.0\times10^{17}(A_*/0.03)^{3/10}(v_{w,8}\,t_{*,6})^{2/5}n_{0,3}^{-3/10}\rm\, cm$,
where $v_{w,8}$ is the wind velocity in units of $10^8\,\rm cm\,s^{-1}$,
$t_{*,6}$ is the lifetime in units of $10^6\,\rm yr$ of the Wolf-Rayet phase presumed 
to have driven the wind, and $n_{0,3}$ is the surrounding interstellar matter (ISM) particle density 
in units of $10^3\,\rm cm^{-3}$.
Hence the two radii are comparable (around $R\sim10^{19}\,$cm)
if, for example, $\eta \sim 0.07$, $A_*=0.03$ and $n_{0,3}$ has a rather low value $\sim0.0014$.
If the prompt gamma-rays were also produced by this jet
then the total energy
would be given by $E_{\rm tot}\approx E_{\gamma}= E_k/\eta 
= 9.8\times10^{51}(A_*/0.03)^{1/2}(\eta/0.07)^{-1/2}\,$erg, comparable to, but somewhat less than, the values 
found above for a single jet with a uniform external medium of density $n \sim
1\,{\rm cm}^{-3}$.
If, on the other hand, the gamma-rays come from a narrow jet, then $E_\gamma$ can be much lower,
and $E_{\rm tot}$ could be dominated by kinetic energy, $E_k=6.9\times 10^{50} (A_*/0.03)^{1/2}(\eta/0.07)^{1/2}\,$erg.

\subsection {The supernova}

The $r$, $i$ and $z$-band observations (Figure~\ref{figlc} right-hand panels)
do not show a break at
the same time as the bluer bands, but rather exhibit at first a flattening optical decay, 
and marked reddening, followed by a steepening again
after about 40 days. 
This is illustrated by the change in color of the optical transient from
$g-i=0.60\pm0.12$ at 14 days post-burst, to
$g-i=1.88\pm0.19$ at 26 days.
We interpret this as being due to the
contribution to the optical light of an underlying supernova that begins  to
dominate the afterglow in the redder bands.
Such supernova ``red humps" have been seen in the
light curves of several long-duration GRBs which have been monitored
sufficiently deeply at late times \citep[eg.][]{galama00,zeh04}.

As stated above, we follow the conventional procedure
of assuming a light curve for the supernova component
based on that of SN1998bw, which accompanied 
the low-redshift GRB\,980425 \citep{galama98,mckenzie99}.
We redshifted and $k$-corrected these  light
curves to produce templates in our observed wavebands
appropriate to $z=0.937$, and faded these by 0.3\,mag,
consistent with the typical GRB-SN ``humps" found by \citet{zeh04}.

When added to the broken power-law
afterglow, this produces quite a reasonable match
to the photometry of the transient.
Thus we find that GRB\,080319B was accompanied
by a supernova a little fainter than the prototype SN1998bw:
whilst an even better match to the photometry would have been achieved with
a supernova model 
having a peak time a little earlier \citep[a stretch factor $<1$
in the language of][]{zeh04}.
This is in slight disagreement with \citet{bloom09} who,
using more preliminary and less complete set of
late-time photometry, concluded that a  supernova
component rather brighter than SN1998bw was required.

\subsection{The host galaxy}

Our second epoch {\em HST} observations revealed that the afterglow, while
still detected, was clearly superimposed upon faint, extended host galaxy emission,
with the transient slightly offset north by about 0.2 arcsec from the center of
this emission \citep{levan08}.
By the third epoch the galaxy clearly dominates and
is revealed  to be a very faint, low surface brightness source extending over
roughly 0.5 arcsec.  This corresponds to a physical size of about 4 kpc
(assuming conventional cosmological parameters) which is quite
typical for GRB hosts \citep{fruchter06}.
The host is not well detected in the WFPC2 images, so the photometry carries
a large uncertainty.
Our best estimates of the host photometry come from the latest
epoch ground-based imaging, with the $g$ band measurement being
corrected to 
remove the residual transient light
by subtracting the flux predicted by our simple model, as described above.
Hence we find (foreground extinction corrected) 
host magnitudes of 
$i({\rm AB})=26.17\pm0.15$,
$r({\rm AB})=26.96\pm0.13$
and $g({\rm AB})=26.81\pm0.14$.

This final photometry, while limited, does allow for a crude fit
to the galaxy SED. In particular the relatively blue $g-r$ color 
($-0.15\pm 0.19$), coupled
with the red $r-i$ ($0.79 \pm 0.20$), is suggestive of a moderately strong Balmer break, implying 
the presence of an older stellar population, in addition to the young population
which produces the blue rest-frame UV color,
and presumably seeded the GRB. 

In Figure~\ref{figsed} we show these photometric points fitted with
a star-forming galaxy template with the following properties:
$M_V=-17.49\pm0.15$ and $M_B=-17.23\pm0.15$ (quoted errors are statistical).
The best fit has an internal extinction $A_V=0$, which although
poorly constrained by our limited photometry, is consistent with
the low extinction seen to the afterglow.
These numbers imply a
star formation rate (SFR) $\approx0.13M_{\odot}$~yr$^{-1}$
and
stellar mass $M\approx1.2\times10^8M_{\odot}$, although observational
and modeling uncertainties make such determinations only accurate
to factors of a few. 
As shown in 
Figure~\ref{fighosts}, this indicates GRB\,080319B has one of the smaller
hosts found to date,  
although note that a small number of
the faintest GRB hosts, which only have photometric detections in one band, are not
included in this figure.
The best-fit  specific star formation rate (SSFR) is therefore
$\Phi=1.1$\,Gyr$^{-1}$ (but with an even greater error bar), which is close to the
average for the sample of $z<1.2$ GRB hosts studied by \citet{svensson10}.

This luminosity corresponds to about $\frac{1}{40}L_*$ 
at the observed redshift \citep[cf.][]{willmer06}.
Such a small galaxy is likely to have low metallicity,
although quantifying this is hard, not least because of the 
small number of data points available for the fit and their
large photometric uncertainties.
Based on the $z\sim0.7$ mass-metallicity relationship of \citet{savaglio05}, the implied metallicity is 
$12+{\rm log}({\rm O/H})=7.9$, or about 20\% of Solar.
However,  this numerical value should be treated with caution for various reasons:
first, the absolute calibration of the mass-metallicity relation 
is difficult, and we note that \citet{savaglio09} using a revised 
calibration based on \citet{kewley08} found metallicities to 
decrease by $\sim0.5$~dex; and second,
in the same paper \citet{savaglio09} show that GRB hosts
with spectroscopically estimated metallicities
scatter quite widely around this relation in any case.

These properties are within the range of other
GRB host galaxies
\citep[e.g.][]{fruchter06,savaglio09},
but place the GRB\,080319B host at the faint end of the available sample.
The location
of the galaxy in the redshift - magnitude plane is shown in Figure~\ref{rz}, which
shows that it is the faintest yet observed by {\em HST} at comparable redshift. 
Similarly, the  model fit would imply an SFR
and a stellar mass at the low end, compared to a sample of other GRB hosts (Figure~\ref{fighosts}).
We caution that a proportion of these redshifts were obtained from
host rather than afterglow spectroscopy, and hence there is some bias
against very faint hosts.  For illustration, hosts without redshift are shown in a separate panel
on the right side of Figure~\ref{rz}.
A particular case in point is that of GRB\,980326 which also exhibited
a ``supernova hump" in its light curve suggestive of a redshift $z\sim1$,
but had a host galaxy with $R>27$ \citep{bloom99}.

\section{Summary}
We have presented a late-time optical and X-ray 
study of the exceptionally
bright GRB\,080319B. These data allow us to decompose the contributions from
afterglow, supernova and underlying host galaxy.  We find that the afterglow of GRB\,080319B 
exhibited an achromatic break in its lightcurve at $\approx 10^6$ seconds, which 
can be interpreted as being due to the relativistic outflow being
initially confined within a jet.
The sharpness of this break is not expected for a simple $R^{-2}$ wind density
profile for the surrounding medium, and may indicate that the jet reaches a termination shock in the pre-existing
wind at about the same radius, $R\sim10^{19}\,$cm.
A simple jet breaking at this time has a total energy, $E_{\rm jet} \gsim 10^{52}\;$erg.
For more complex jet structures in which the gamma-ray and late
afterglow arise from different components, such as the two component jet
model of Racusin et al. (2008), the total jet energy can be smaller.

In addition GRB\,080319B was associated
with a bright supernova, slightly fainter in luminosity than the  prototype SN1998bw.
Such supernovae, inferred from ``red humps" in their light curves,
have been found to accompany several other GRBs at similar
redshifts \citep[e.g.][]{zeh04}. Indeed, apart from the few (generally low
luminosity) bursts with spectroscopically confirmed supernova components, 
the data-set for  GRB\,080319B provides one of the most compelling examples.

Finally, we have detected a small host galaxy under the position of the GRB, 
which is fainter than other GRB hosts observed so far at comparable redshifts.
This is likely to indicate a low-metallicity environment, and one
might speculate that this could be related to the extreme properties of the
burst.  However it is also notable that most of the weakest GRBs known
(particularly GRB\,980425 and GRB\,060218)
also have occurred in small, low-metallicity hosts and been
accompanied by energetic type Ibc supernovae \citep[e.g.][]{stanek06,wiersema07}.

\acknowledgments

We acknowledge useful discussions with Daniele Malesani and Ralph Wijers.

We are grateful to Matt Mountain for awarding directors discretionary time on HST to 
observe GRB\,080319B under program GO/DD 11513 (PI: Tanvir). 
Based on observations made with the NASA/ESA {\em Hubble Space Telescope}, obtained at the Space Telescope Science Institute, which is operated by the Association of Universities for Research in Astronomy, Inc., under NASA contract NAS 5-26555.

Based on observations obtained at the Gemini Observatory, which is operated by the
Association of Universities for Research in Astronomy, Inc., under a cooperative agreement
with the NSF on behalf of the Gemini partnership: the National Science Foundation (United
States), the Science and Technology Facilities Council (United Kingdom), the
National Research Council (Canada), CONICYT (Chile), the Australian Research Council
(Australia), MinistŽrio da Cincia e Tecnologia (Brazil) and SECYT (Argentina)

This publication has made use of
data obtained with the {\em Chandra X-ray Observatory}, under program ID
09500789.
(observations IDs \dataset[ADS/
Sa.CXO#Obs/09134]{9134} and \dataset[ADS/Sa.CXO#Obs/09134]{9135})

Based on observations made with ESO Telescopes at the La
Silla or Paranal Observatories under programme ID 081.D-0853

The DARK Cosmology Centre is funded by the
Danish National Research Foundation.

We particularly thank the staff of the VLT and Gemini for their
efforts in obtaining the optical data, and those at CXC for their assistance in
scheduling the {\em Chandra} observations. We also gratefully acknowledge the work of the wider \swift\
team that makes this research possible.
NRT, ER and AJL are supported by 
STFC.  JG gratefully
acknowledges a Royal Society Wolfson Research Merit Award. 
AJvdH is supported by an appointment to the NASA
Postdoctoral Program at the MSFC, administered by ORAU through a
contract with NASA.


%% in parentheses, after the keyword, but they will not be verified.

{\it Facilities:}  \facility{HST (WFPC2)}, \facility{CXO (ACIS)}, \facility{Gemini-North(GMOS)},
\facility{VLT(FORS1)}.

\clearpage

%% the file, uncomment the \plotone, \plottwo, and
%% If you need a layout that cannot be achieved with \plotone or
%% \plottwo, you can invoke the graphicx package directly with the02
%% \includegraphics command or use \plotfiddle. For more information,
%% please see the tutorial on "Using Electronic Art with AASTeX" in the
%% documentation section at the AASTeX Web site,
%% http://www.journals.uchicago.edu/AAS/AASTeX.
%%
%% The examples below also include sample markup for submission of
%% supplemental electronic materials. As always, be sure to check
%% the instructions to authors for the journal you are submitting to
%% for specific submissions guidelines as they vary from

%% This example uses \plotone to include an EPS file scaled to
%% 80% of its natural size with \epsscale. Its caption
%% has been written to indicate that additional figure parts will be
%% available in the electronic journal.

\begin{figure}
\includegraphics[angle=0,scale=0.945]{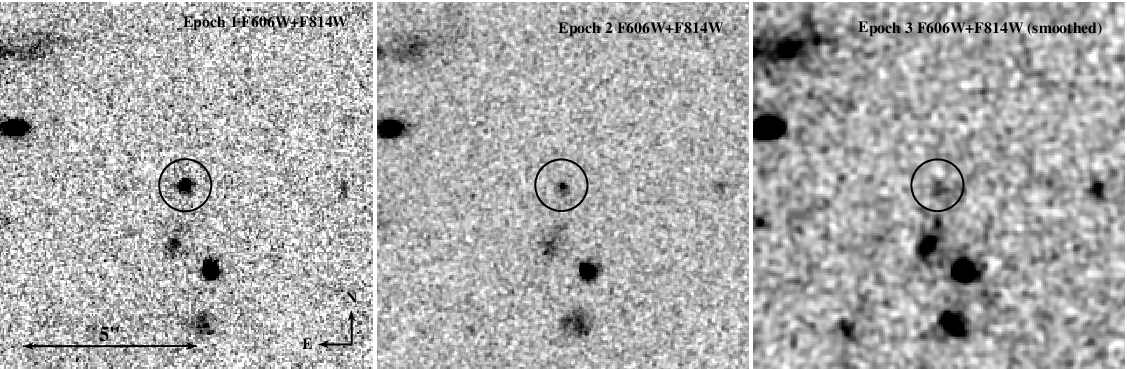}
\caption{
{\it HST}/WFPC2 images (F606W and F814W combined) at each epoch, as
labeled.  The circle, whose radius is arbitrary, is centered at the position
of the afterglow. The phases are afterglow, supernova and host
dominated, respectively.  Note that the final panel has
been smoothed with a Gaussian kernel to bring out the
faint host galaxy light.
\label{figmosaic}}
\end{figure}

\begin{figure}
%\epsscale{.80}
%\plotone{080319B_lc.ps}
\includegraphics[angle=270,scale=0.7]{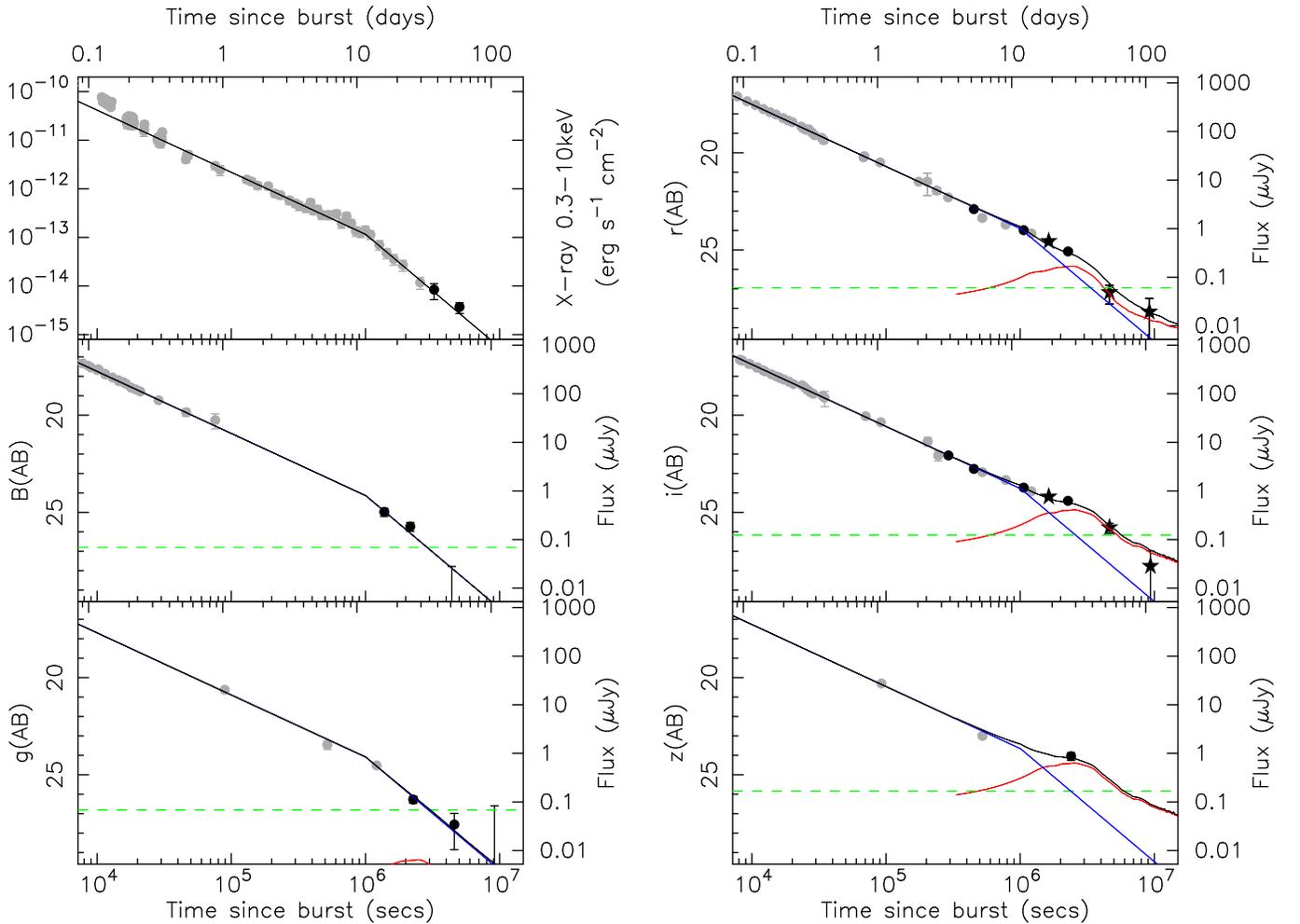}
\caption{
Late time photometry of GRB\,080319B, with bold symbols indicating
observations reported here, and light symbols being data-points from
the literature \citep[][and references therein]{racusin08,bloom09}.  
Photometry has been corrected for foreground
extinction, and error bars are $1\sigma$, although in many cases these
are smaller than the symbol size.
The green dashed lines are the
estimated magnitudes of the host galaxy which have been subtracted
from the ground-based (filled circles) 
data.  In the case of {\em HST} images the point
source photometry (filled stars)
is done on a scale smaller than the host, and the
contribution within the aperture estimated from the latest time images.
The blue line is the model afterglow, and the red line is the model
supernova light curve, as described in the text.  The black line is their
sum.
\label{figlc}}
\end{figure}

\begin{figure}
\includegraphics[angle=0,scale=0.4]{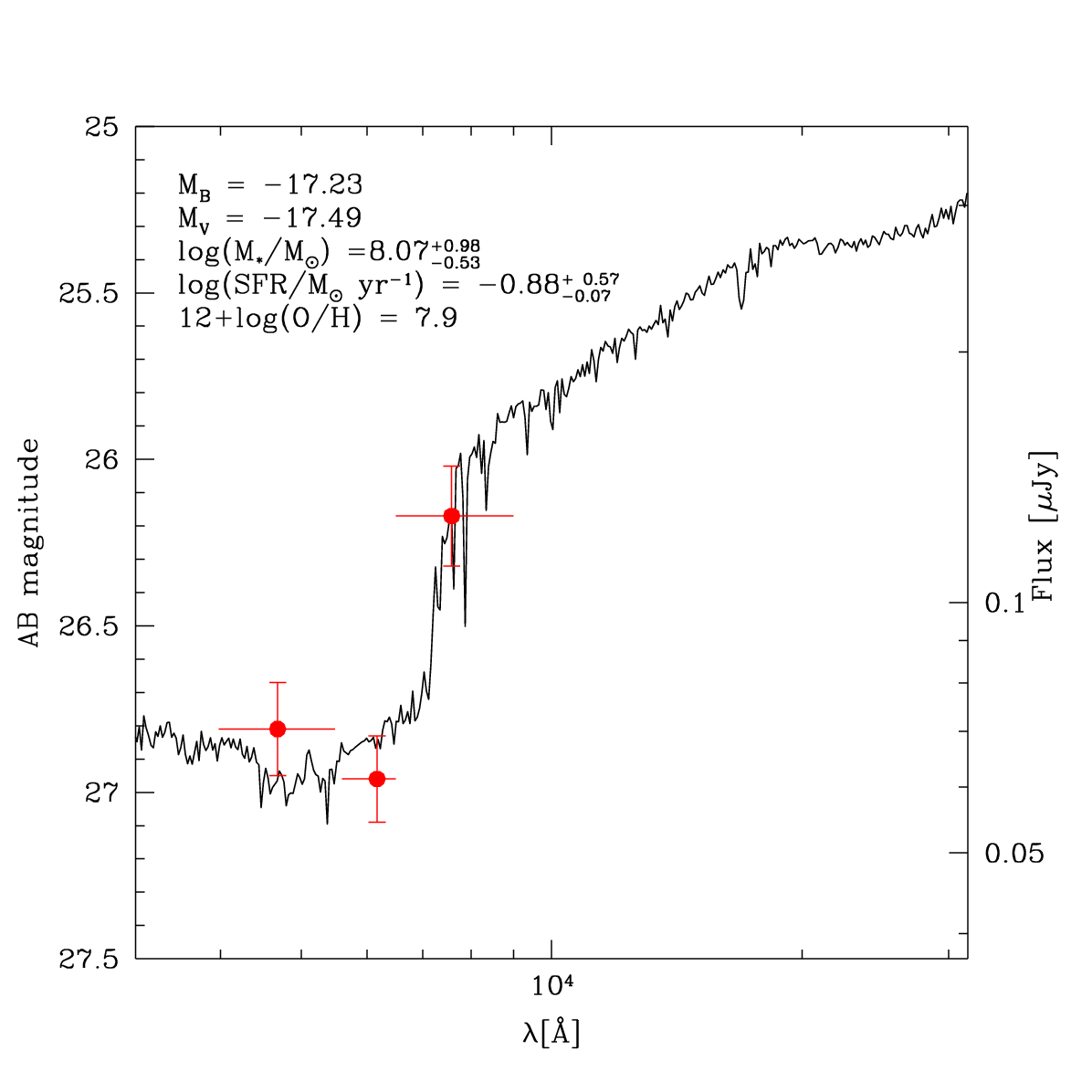}
\caption{
A model spectrum fitted to the three host photometry points.
The best fit model is a young star forming galaxy \citep{BruzCharl03}. Physical parameters are derived from
the restframe SED:  star formation rates (SFRs) are derived from the $U$-band luminosity \citep{cram98};
stellar mass is estimated from the $K$-band luminosity with a color correction to the
mass-to-light ratio \citep{mannucci05}. The $\chi^{2}_{\nu}$ of the fit is a very acceptable $0.89$.
\label{figsed}}
\end{figure}

\begin{figure}
\includegraphics[angle=0,scale=0.45]{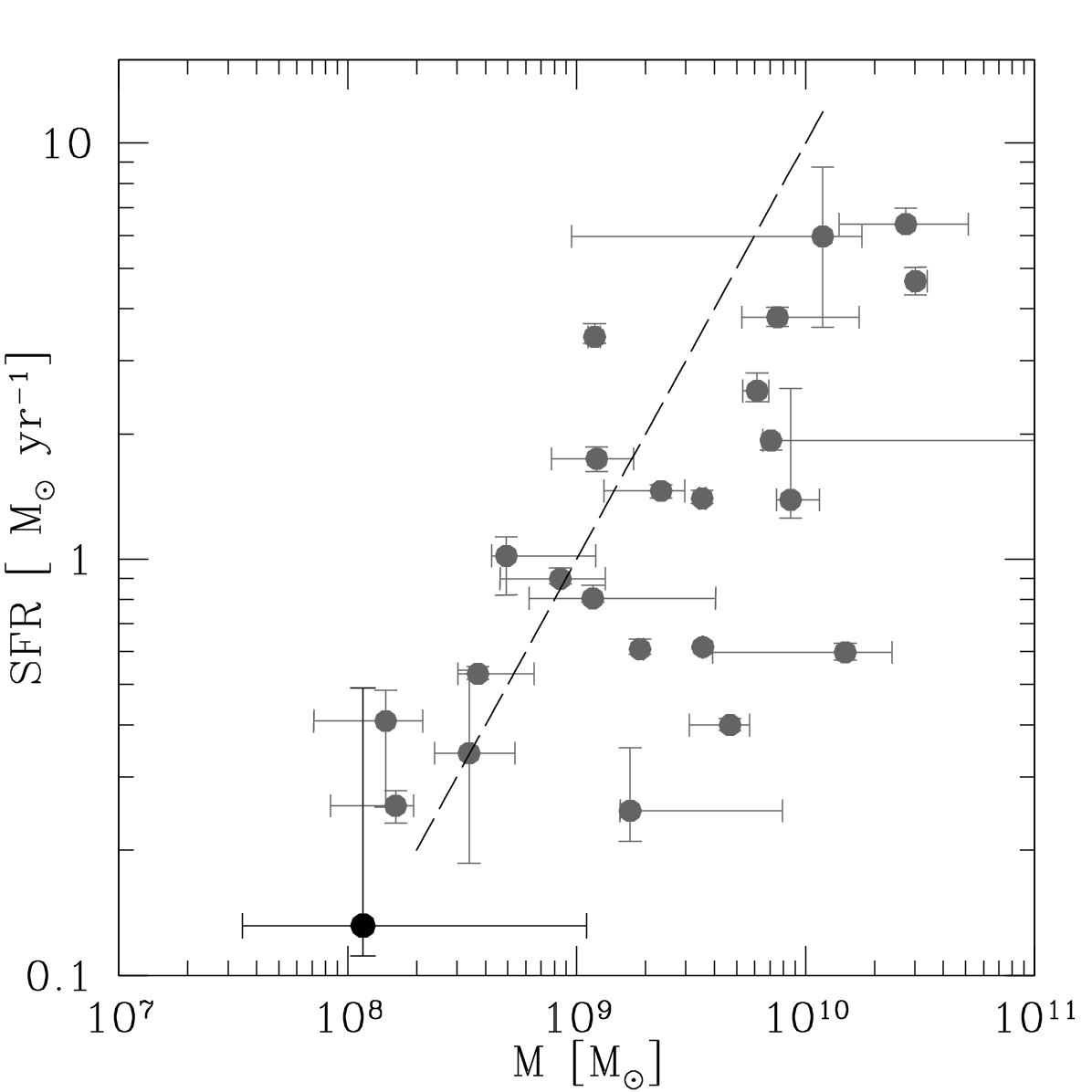}%080319_M_sfr.eps}
\caption{
Plot of star formation rate versus stellar mass for a sample of
GRB hosts, inferred from template fitting to their photometric SEDs.
The host of GRB\,080319B, shown by a  bold symbol, is at the 
lower mass  and star-formation rate end of the distribution.
Within the estimated errors, its specific star formation rate seems about average
for the sample as a whole.
The dashed line shows a locus of constant specific star formation rate, illustrating
the typically higher SSFRs for lower stellar mass galaxies.
The host galaxy sample is that used by 
\citet{svensson10}
and includes all GRB hosts with $z<1.2$ and at least two
photometric detections.
Parameters determined by SED fits to the photometry, as detailed
in the caption to Fig.~\ref{figsed}.
The error bars represent 68\% confidence determined by the model fits,
and illustrate that while neither the stellar mass nor the star-formation rate is
tightly constrained by this method, the latter is rather better determined 
thanks to the detection of flux below the Balmer break, which is dominated
by the young stellar component.
\label{fighosts}}
\end{figure}

\begin{figure}
\includegraphics[angle=0,scale=0.4]{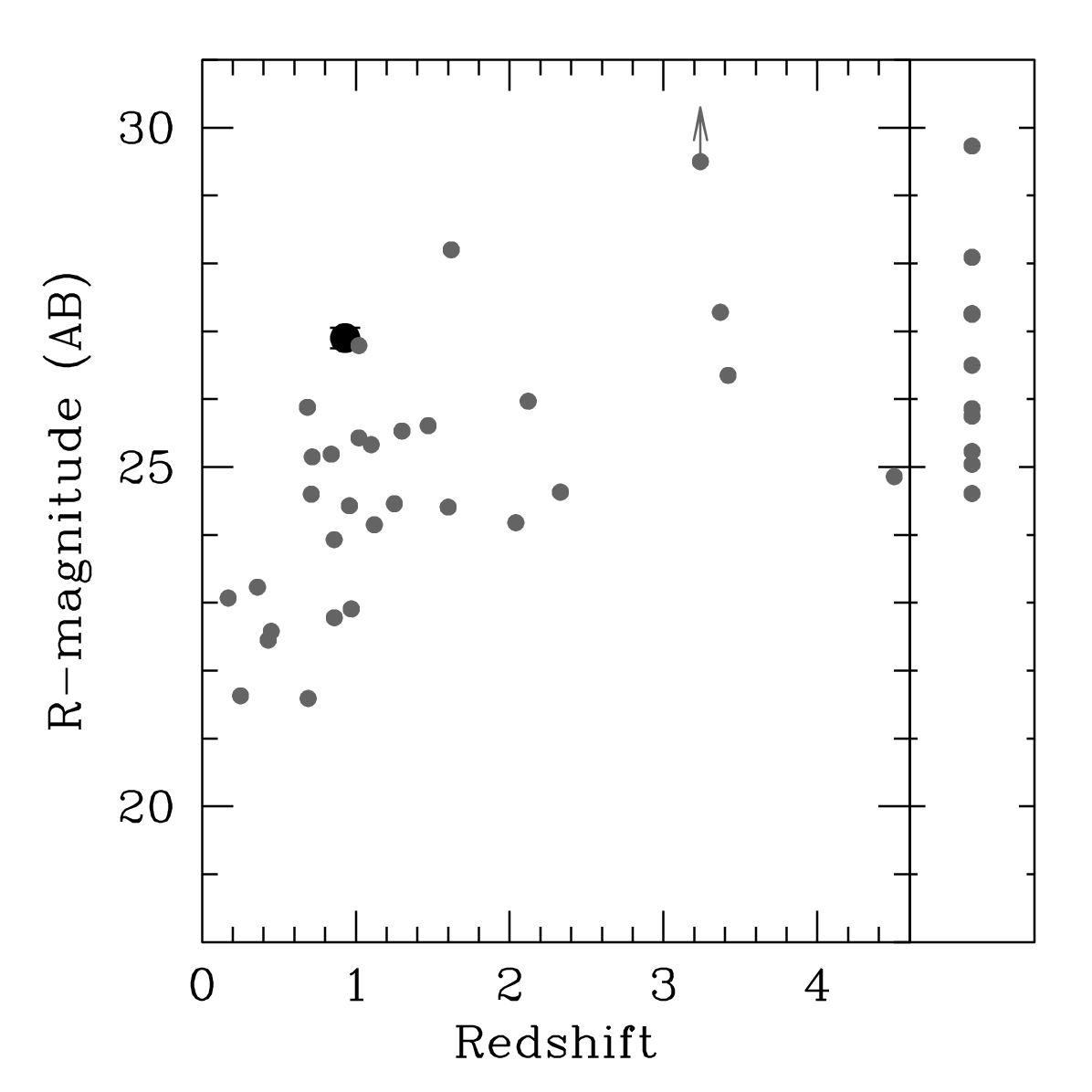}
\caption{
The $R$(AB) magnitude of the host of GRB\,080319B compared to
a sample of other GRB hosts observed by {\em HST} as a function of 
redshift.  Those without redshifts are shown in the right hand panel.
Clearly the GRB\,080319B host is faint, even by the standards of
other GRB hosts.
\label{rz}}
\end{figure}

\clearpage

\begin{table}
\begin{center}
\caption{Log of the late-time observations reported here. This photometry 
%is CTE corrected, but 
is not corrected for extinction, but the small aperture {\it HST} photometry
has been aperture and CTE corrected. 
Note that these
fluxes include both transient light and host light within the apertures, whereas in
creating Figure 2 the host contribution was modeled and removed, as described
in the text.\label{tbl-1}  }
\begin{tabular}{lllrllr}
\tableline\tableline
Time post- & Telescope/ & Filter & Exposure  & Flux  & Error  & Aperture\\
burst (days) & camera &  &  time (s) & ($\mu$Jy) &   & (diameter arcsec)\\
%burst (days) & camera &  &  time (s) & AB Magnitude &   & (diameter arcsec)\\
\tableline
%
% Trigger at Mar19 6.13 UT (MJD 54544.25903)
%

3.36 & Gemini-N/GMOS & $r$ & $5\times 200$  & 4.33 & 0.08 & 2.0\\  

3.38 & Gemini-N/GMOS & $i$ & $5\times200$  & 5.40 & 0.10 & 2.0\\  

5.22 & Gemini-N/GMOS & $r$ & $5\times100$  & 2.51 & 0.09 & 2.0 \\  

5.23 & Gemini-N/GMOS & $i$ & $5\times100$  & 2.91 & 0.11 & 2.0\\  

12.3 & Gemini-N/GMOS & $r$ & $5\times100$  & 0.96 & 0.05 & 2.0 \\  

12.3 & Gemini-N/GMOS & $i$ & $5\times100$  & 1.27 & 0.07 & 2.0 \\  

26.3$^1$ & Gemini-N/GMOS & $g$ & $6\times180$  &  0.174& 0.013 & 1.5\\  

26.3$^1$ & Gemini-N/GMOS & $r$ &  $6\times180$ & 0.39 & 0.03 & 1.5 \\  

26.3$^1$ & Gemini-N/GMOS & $i$ & $6\times180$  & 0.74 & 0.03 & 1.5 \\  

27.7$^1$ & Gemini-N/GMOS & $z$ &$6\times180$  & 1.09 & 0.16 & 1.5 \\  

53.2 & Gemini-N/GMOS & $g$ &$9\times 300$  & 0.099 & 0.020 & 1.5 \\  

106.1 & Gemini-N/GMOS & $g$ & $9\times350$ & 0.072 & 0.007 & 1.5 \\

319.3 & Gemini-N/GMOS & $r$ & $10\times450$ & 0.062 & 0.009 & 1.5 \\

463.1 & Gemini-N/GMOS & $i$ & $10\times360$ & 0.125 & 0.019 & 1.5 \\

16.0 & VLT/FORS1 & $B$ & $6\times300$  & 0.49 & 0.08 & 1.5\\ 

25.0 & VLT/FORS1 & $B$ & $6\times300$  & 0.24 & 0.04 & 1.5\\ 

51.0 & VLT/FORS2 & $B$ & $18\times300$  & 0.071 & 0.023 & 1.5\\ 

18.9$^2$ & {\em HST}/WFPC2 & F606W & $8\times400$  & 0.55 & 0.01 & 0.2 \\ 

19.1$^2$ & {\em HST}/WFPC2 & F814W & $8\times400$  & 0.81 & 0.02 & 0.2 \\ 

53.4$^3$ & {\em HST}/WFPC2 & F814W & $8\times400$  & 0.200 & 0.023 & 0.2 \\ 

53.6$^3$ & {\em HST}/WFPC2 & F606W & $8\times400$  & 0.066 & 0.006 & 0.2 \\ 

106.4 & {\em HST}/WFPC2 & F606W & $8\times400$  & 0.033 & 0.009 & 0.2\\ 
 &  &  &   & 0.065 & 0.028 & 1.0\\ 

108.3 & {\em HST}/WFPC2 & F814W & $8\times400$  & 0.046 & 0.027 & 0.2\\ 
 &  &  &   & 0.219 & 0.045 & 1.0\\ 

\tableline
\end{tabular}
% Any table notes must follow the \end{tabular} command.
\tablenotetext{1}{Independent reduction of these data already reported in 
%Tanvir et al. (2008b), Bloom et al. (2009)}
\citet{tanvirgcn2}, \citet{bloom09}}
\tablenotetext{2}{Provisional photometry already reported in \citet{tanvirgcn1}, \citet{racusin08}}
\tablenotetext{3}{Provisional photometry already reported in \citet{levan08}}
%generated with the \LaTeX\ table environment}
%\tablenotetext{b}{Yet another sample footnote for table~\ref{tbl-2}}
%\tablenotetext{c}{Another sample footnote for table~\ref{tbl-2}}
%\tablecomments{We can also attach a long-ish paragraph of explanatory
%material to a table.}
\end{center}
\end{table}

%% If the table is more than one page long, the width of the table can vary
%% from page to page when the default \tablewidth is used, as below.  The
%% individual table widths for each page will be written to the log file; a
%% maximum tablewidth for the table can be computed from these values.
%% The \tablewidth argument can then be reset and the file reprocessed, so
%% that the table is of uniform width throughout. Try getting the widths
%% from the log file and changing the \tablewidth parameter to see how
%% adjusting this value affects table formatting.

%% The \dataset{} macro has also been applied to a few of the objects to
%% show how many observations can be tagged in a table.

%\clearpage

%% Tables may also be prepared as separate files. See the accompanying
%% sample file table.tex for an example of an external table file.
%% To include an external file in your main document, use the \input
%% command. Uncomment the line below to include table.tex in this
%% sample file. (Note that you will need to comment out the \documentclass,
%% \begin{document}, and \end{document} commands from table.tex if you want
%% to include it in this document.)


%% The following command ends your manuscript. LaTeX will ignore any text


\begin{thebibliography}{}
\bibitem[Adelman-McCarthy et al.(2007)]{sdss}  Adelman-McCarthy, J.~K., et al.\ 2007, \apjs, 172, 634 
\bibitem[Akerlof et al.(1999)]{akerlof99} Akerlof, C., et al. 1999, \nat, 398, 400 
\bibitem[Bloom et al.(1999)]{bloom99}  Bloom, J.~S., et al., 1999, \nat, 401, 453 
\bibitem[Bloom et al.(2009)]{bloom09} Bloom, J.~S., et al. 2009, \apj, 691, 723
\bibitem[Bruzual \& Charlot(2003)]{BruzCharl03} Bruzual, A. G., \& Charlot, S.\ 2003, \mnras, 344, 1000 
\bibitem[Cool et al.(2008)]{cool08} Cool, R.~J., et al.\ 2008,  GRB Coordinates Network, 7465, 
\bibitem[Cram et al.(1998)]{cram98} Cram, L., Hopkins, A., Mobasher, B., \& Rowan-Robinson, M.\ 1998, \apj, 507, 155 
\bibitem[Curran, van der Horst, \& Wijers (2008)]{curran08} Curran, P.~A., van der Horst, A.~J., \& Wijers, R.~A.~M.~J., 2008, MNRAS, 386, 859 
\bibitem[Fruchter  \& Hook(2002)]{fruchter02} Fruchter, A.~S., \& Hook, R.~N.\ 2002, \pasp, 114, 144 
\bibitem[Fruchter et al.(2006)]{fruchter06} Fruchter, A.~S., et  al.\ 2006, \nat, 441, 463 
\bibitem[Galama et al.(1998)]{galama98} Galama, T.~J., et al. 1998, \nat, 395, 670
\bibitem[Galama et al.(2000)]{galama00} Galama, T.~J., et al. 2000, \apj, 536, 185
\bibitem[Granot, K{\"o}nigl, \& Piran (2006)]{granot06} Granot, J., K{\"o}nigl, A., \& Piran, T., 2006, MNRAS, 370, 1946 
\bibitem[Haislip et al.(2006)]{haislip06} Haislip, J.~B., et al. 2006, \nat, 440, 181 
\bibitem[Kewley \& Ellison(2008)]{kewley08} Kewley, L.~J., \& Ellison, S.~L.\ 2008, \apj, 681, 1183 
\bibitem[Kraft et al.(1991)]{kraft91} Kraft, R.~P., Burrows, D.~N., \& Nousek, J.~A.\ 1991, \apj, 374, 344
\bibitem[Kumar \& Narayan(2009)]{Kumar09} Kumar, P., Narayan, R., 2009, \mnras, 395, 472 
\bibitem[Kumar \& Panaitescu(2000)]{kumar00} Kumar, P., \& Panaitescu, A.\ 2000, \apj, 541, L51
\bibitem[Kumar \& Panaitescu(2008)]{kumar08} Kumar, P., \& Panaitescu, A.\ 2008, \mnras, 391, L19
\bibitem[Levan et al.(2008)]{levan08} Levan, A.~J., Tanvir, N.~R., \& Fruchter, A.~S. 2008, GRB Coordinates Network, 7710
\bibitem[Mannucci et al.(2005)]{mannucci05} Mannucci, F., Della Valle, M., Panagia, N., Cappellaro, E., Cresci, G., Maiolino, R., Petrosian, A., \& Turatto, M.\ 2005, \aap, 433, 807 
\bibitem[McKenzie et al.(1999)]{mckenzie99} McKenzie, E.~H. \& Schaefer, B.~E.\ 1999, \pasp, 111, 964
\bibitem[Nakar \& Granot (2007)]{Nakar07} Nakar, E. \& Granot, J., 2007, \mnras, 380, 1744 
\bibitem[Nousek et al. (2006)]{nousek06} Nousek, J.~A., et al., 2006, \apj, 642, 389 
\bibitem[Panaitescu \& Kumar (2000)]{Panaitescu2000} Panaitescu, A. \& Kumar, P., 2000, \apj, 543, 66 
\bibitem[Pandey et al. (2009)]{pandey09}  Pandey, S.~B., et al., 2009, A\&A, 504, 45 
\bibitem[Pe'er \& Wijers (2006)]{Peer2006} Pe'er A. \& Wijers R.~A.~M.~J., 2006, \apj, 643, 1036 
\bibitem[Pei(1992)]{pei92} Pei, Y.~C.\ 1992, \apj, 395, 130
\bibitem[Peng, K{\"o}nigl \& Granot (2005)]{peng05} Peng, F., K{\"o}nigl, A., \& Granot, J.\ 2005, \apj, 626, 966 
\bibitem[Piran et al.(2009)]{piran09} Piran, T., Sari, R., \& Zou, Y.-C.\ 2009, \mnras, 393, 1107 
\bibitem[Price et al.(2002)]{Price02} Price, P.~A., et al., 2002, ApJ, 572, L51 
\bibitem[Racusin et al.(2008)]{racusin08} Racusin, J.~L., et al. 2008, \nat, 455, 183
\bibitem[Rhoads(1999)]{rhoads99} Rhoads, J.~E.\ 1999, \apj, 525, 737 
\bibitem[Sari, Piran, \& Halpern (1999)]{sari99} Sari R., Piran T., \& Halpern, J.~P., 1999, ApJ, 519, L17 
\bibitem[Savaglio et al.(2009)]{savaglio09} Savaglio, S.,  Glazebrook, K., \& Le Borgne, D.\ 2009, \apj, 691, 182 
\bibitem[Savaglio et al.(2005)]{savaglio05} Savaglio, S., et al. 2005, \apj, 635, 260 
\bibitem[Schlegel et al.(1998)]{schlegel98} Schlegel, D.~J.,  Finkbeiner, D.~P., \& Davis, M.\ 1998, \apj, 500, 525 
\bibitem[Stanek et al. (2006)]{stanek06} Stanek, K.~Z., et al., 2006, Acta Astron., 56, 333 
\bibitem[Svensson et al.(2010)]{svensson10} Svensson, K.~M., Levan, A.~J., Tanvir, N.~R.,  Fruchter, A.~S. \& Strolger, L.-G.\ 2010, \mnras, 405, 57
\bibitem[Tanvir et al.(2008a)]{tanvirgcn1} Tanvir, N.~R., Levan,  A.~J., Fruchter, A.~S., Graham, J., Wiersema, K., \& Rol, E.\ 2008a, GRB Coordinates Network, 7569
\bibitem[Tanvir et al.(2008b)]{tanvirgcn2} Tanvir, N.~R., Perley, D.~A., Levan, A.~J., Bloom, J.~S., Fruchter, A.~S., \& Rol, E.\ 2008b, GRB Coordinates Network, 7621
\bibitem[Vreeswijk et al. (2008)]{vreeswijk08} Vreeswijk, P., et al. 2008, GRB Coordinates Network, 7444
\bibitem[Wiersema et al.(2007)]{wiersema07} Wiersema, K., et al., 2007, A\&A, 464, 529 
\bibitem[Willmer et al.(2006)]{willmer06} Willmer, C.~N.~A., et al., 2006, ApJ, 647, 853 
\bibitem[Wozniak et al. (2009)]{wozniak09} Wozniak, P.~R., Vestrand, W.~T., Panaitescu,
A.~D., Wren, J.~A., Davis, H.~R., White, R.~R., 2009, \apj, 691, 495
\bibitem[Zeh et al.(2004)]{zeh04} Zeh, A., Klose, S., \& Hartmann, D.~H. 2004, \apj, 609, 952 
\bibitem[Zou, Piran, \& Sari(2009)]{Zou09} Zou, Y.-C., Piran, T., \& Sari, R., 2009, \apj, 692, L92 
\end{thebibliography}
\end{document}